\documentstyle[aps,pre,eqsecnum,multicol]{revtex}
\begin{document}
\draft
\title{Universal long-time properties of Lagrangian statistics in 
the Batchelor regime and their application to the passive scalar problem}
\author{E.~Balkovsky and A.~Fouxon}
\address{Department of Physics of Complex Systems, Weizmann Institute
of Science, Rehovot, 76100, Israel}
\date{\today}
\maketitle

\begin{abstract}
We consider the transport of dynamically passive quantities in the
Batchelor regime of a smooth in space velocity field. For the case of arbitrary
temporal correlations of the velocity, we formulate the statistics of relevant
characteristics of Lagrangian motion. This allows us to generalize many
results obtained previously for strain $\delta$ correlated in time, thus
answering a question about the universality of these results.
\end{abstract}

\pacs{PACS numbers 47.27.Eq, 05.40.-a, 47.10.+g}

\begin{multicols}{2}

\section*{Introduction}

The problem of passive scalar transport by turbulent flows has received much
attention lately. The progress achieved has been made possible mainly by
the introduction of the Kraichnan model \cite{Kra68}. Within the model the
turbulent velocity statistics is believed to be Gaussian, scale invariant in
space and $\delta$ correlated in time, which allows one to write closed
equations on the correlation functions of the scalar. Such a velocity has only
a few rough features in common with realistic flows, which are intermittent
and have a finite scale-dependent correlation time, contrary to what is
assumed in the model. Nevertheless, it seems that many interesting properties
of the statistics are inherent in the dynamics, rather than due to the
intermittency of the velocity statistics itself. Unusually for the turbulence
theory numerous results have been obtained analytically using the Kraichnan
model.

Having reached an understanding of this model, it is then natural to
generalize its results, passing to more realistic flows. However, due to the
complicated interplay between spatial and temporal properties of the velocity,
one encounters various difficulties in introducing a meaningful velocity field
with a finite correlation time. The only case where this was easily done is
the so called Batchelor regime \cite{Bat59}, where the spatial structure of the
velocity is rather simple, and therefore one can separate space and time
dependencies. It appears in the limit of large Prandtl numbers, which is the
ratio of the fluid viscosity to the diffusivity of the transported quantity.
In studying advection below the viscous length, the correlation functions of
the velocity are smooth functions of space, which allows one to introduce an
effective description with $v_\alpha=\sigma_{\alpha_\beta}(t)r_\beta$
\cite{Bat59}. In this way time and space become completely separated. 

The Batchelor limit is well studied if $\sigma$ has a zero correlation time and
its statistics is Gaussian
\cite{SS94,CFKL95,BGK97,Son98,CFK98,CKV97,Fou98,KLS98,CGK94,BFLL99,Kra74};
that is, in the framework of the Kraichnan model. Certain results have been
derived for arbitrary statistics of $\sigma$
\cite{Kra68,Bat59,SS94,CFKL95,Kra74}.

Our aim here is to investigate the degree of universality of the passive
scalar statistics for arbitrary temporal correlations of the velocity. We
utilize the close relation between the statistics of Lagrangian trajectories
in a turbulent flow and the statistics of the passive scalar. Therefore, it
seems reasonable to separate the problems, first investigating the Lagrangian
motion and next applying the results to particular problems. For the Batchelor
regime only a few degrees of freedom characterize the Lagrangian dynamics,
which makes the problem solvable.

The plan of this paper is as follows. First we pass to the comoving reference
frame in the equation for a passive scalar, which allows us to consider the
Lagrangian mapping as an affine transformation, characterized by a random
matrix. After its probability distribution  function (PDF) is found, we
consider several particular examples of the scalar statistics both for the
decaying and forced turbulence.  We show that the statistics of the scalar
can be found by integration of the distribution function with a kernel,
depending on the problem in question.

\section{General relations}

Advection of a passive scalar $\vartheta$ by incompressible velocity field
${\bbox v}$ is described by the equation
\begin{eqnarray}&&
\partial_t \vartheta+({\bbox v},\nabla)\vartheta-\kappa\nabla^2 \vartheta=0\,,
\label{first}\end{eqnarray}
where $\kappa$ is the molecular diffusivity. We shall be interested in the
limit of small but finite $\kappa$. In the case of continuous injection of the
scalar, one should add a source $\phi(t,{\bbox r})$ into the right-hand side of
Eq. (\ref{first}).

Let us consider a blob of the scalar having a size $L$ much smaller than the
viscous length of the velocity. The variation of the velocity on the scale of
the blob is much smaller than the large homogeneous velocity transferring the
blob as a whole. To account for a slow variation of the form of the blob due to
relative motion of the particles, it is natural to pass to the reference frame
moving with the velocity of a particle within the blob \cite{TL,MY}. Since the
velocity is a smooth function on the scale of the blob, it can be expanded in
a Taylor series, thus leading to the equation
\begin{eqnarray}&&
\partial_t\vartheta+\sigma_{\alpha\beta}r_\beta\nabla_{\alpha}\vartheta
-\kappa\nabla^2\vartheta=0\,.
\label{pseq}\end{eqnarray}
Here $\sigma_{\alpha\beta}(t)$ is the matrix of the velocity derivatives taken
at the chosen Lagrangian point. Incompressibility implies
$\sigma_{\alpha\alpha}=0$. For turbulent flows $\sigma$  should be regarded as
a random matrix, having a finite correlation time $\tau$, which is the
Lagrangian correlation time of the velocity.

The complete information about the Lagrangian flow, defined by
$\partial_t {\bbox R}=\sigma{\bbox R}$, is contained in the matrix
$W$, satisfying 
$$\partial_t W=\sigma W\,,\quad W(0)=1\,.$$
This generates an affine transformation of space points, so that a vector
${\bbox R}$ transforms as ${\bbox R}(t)=W(t){\bbox R}(0)$. The volume
conservation guarantees $\det W=1$. The motion of the particles of the scalar
differs from that of the space points due to the nonzero diffusivity. To
investigate this motion we introduce the ''inertia tensor'' of the blob
\cite{TL},
\begin{eqnarray}&&
I_{\alpha\beta}=\frac{1}{2N}
\int d{\bbox r}\,r_\alpha r_\beta\,\vartheta(t,{\bbox r})\,,
\label{I1}\end{eqnarray}
where $N=\int\!d{\bbox r}\,\vartheta(t,{\bbox r})$ is the number of
particles of the scalar. It is easy to check that $N$ is conserved by the full
equation (\ref{pseq}). It turns out that $I$ contains all the necessary
information, and will appear in the following sections as the result of formal
calculations.  The tensor $I$ satisfies the closed dynamical equation
\begin{eqnarray}&&
\partial_t I=\kappa+\sigma I+I\sigma^T\,.
\label{I2}\end{eqnarray}
The initial condition depends on the form of the initial blob, generally
$I_{\alpha\beta}\sim L^2$. We shall see that for problems with spatial
isotropy it is enough to consider $I_{\alpha\beta}(0)=L^2\delta_{\alpha\beta}$.
One can check that $I$ can be expressed via $W$ in a way that is nonlocal in
time.
Since $I$ is symmetric and incorporates the diffusion, instead of
working with $W$ it will be more convenient for us to work directly with
Eq. (\ref{I2}).

The dynamics of the symmetric matrix $I$ can be separated into the nontrivial
essential dynamics of its eigenvalues and the trivial dynamics of the angular
degrees of freedom. It is thus natural to reformulate the dynamics for the
eigenvalues directly, excluding irrelevant angular degrees of freedom. For the
case that is $\delta$ correlated in time, this can be done exactly, resulting
in the Calogero-Sutherland model \cite{BGK97}. We shall show that for a finite
correlation time of $\sigma$ the angular degrees of freedom can also be
effectively excluded. The reason for this is that only the large-time dynamics
of the eigenvalues is important for our purposes so that in many respects (but
not all) the matrix $\sigma$ appears to be $\delta$ correlated in time.

Before we proceed with the derivation, it is useful to understand
qualitatively the typical dynamics of a blob. If the amplitude of the velocity
fluctuations is large enough (the precise condition will be formulated below),
the term $\kappa$ on the right-hand side of Eq. (\ref{I2}) can be disregarded
during the initial stage of the evolution. Then one can make sure that $I$
coincides with $L^2WW^{T}$.  According to the Oseledets theorem \cite{GSO87},
at large enough times the logarithms of the eigenvalues of the latter matrix
are asymptotically equal to $2\lambda_i t$. The Lyapunov exponents $\lambda_1$,
$\dots$, $\lambda_d$ do not depend on a particular  realization of $\sigma$;
hence they are important characteristics of the system.

The above implies that the directions corresponding to positive and negative
Lyapunov exponents will grow or decrease correspondingly, and that the blob
will become an ellipsoid with the length of its main axis changing as
$\exp(\lambda_i t)$. The orientation of the ellipsoid can be arbitrary.
The smallest dimension will decrease exponentially with the rate $\lambda_d$,
until at $t \approx |\lambda_d|^{-1}\ln\left(L^2|\lambda_d|/\kappa\right)$ it
reaches a scale $r_{\rm dif}=\sqrt{\kappa/|\lambda_d|}$, where the
diffusive spreading of particles makes further contraction impossible.
Later, the smallest direction will fluctuate around $r_{\rm dif}$. This will
not affect other directions, that will continue to change according to their
Lyapunov exponents.  In order to have a wide separation of the scales
$L$ and $r_{\rm dif}$, one should require a large value of the Peclet number
\begin{equation}
{\rm Pe}\equiv L/r_{\rm dif}=L\sqrt{|\lambda_d|/\kappa}\,.
\end{equation}
This ensures that the time needed to reach the diffusion scale is large, so
that the above arguments are valid.

Apart from the typical event described here, we shall also need the
distribution of all outcomes. This is the aim of Sec. \ref{Stwo}. Although it
is not difficult to work with the arbitrary dimensionality of space,
we shall consider the physical dimensionalities $d=2$ and $3$ only.

\section{Statistics of $I$}\label{Stwo}

To separate the angular degrees of freedom from the radial ones, it is natural
to represent $I$ as follows:
\begin{equation}
I=R^T \Lambda R\,.
\label{transf}
\end{equation}
Here $R$ is an orthogonal matrix composed of the eigenvectors of $I$, and
$\Lambda$ is a diagonal matrix with the eigenvalues $e^{2\rho_1},\ldots
e^{2\rho_d}$ along the diagonal (we believe that the eigenvalues are ordered,
so that $\rho_1\geq\rho_2\geq\ldots\geq\rho_d$). Equation (\ref{I2}) becomes
\begin{eqnarray}&&
\partial_t \rho_i=\tilde\sigma_{ii}+\frac{\kappa}{2}\exp(-2\rho_i),\quad\quad
\tilde\sigma=R\sigma R^T\,,\label{prt}\\&&
\partial_tR=\Omega R\,,\quad\quad
\Omega_{ij}=\frac{e^{2\rho_i}\tilde\sigma_{ji}+e^{2\rho_j}\tilde\sigma_{ij}}
{e^{2\rho_i}-e^{2\rho_j}}\,.
\label{efd}\end{eqnarray}
We do not assume a summation over the repeating indices in Eqs. (\ref{prt})
and (\ref{efd}). This system of equations is not very useful for analyzing the
general case. However, one can notice that if during the evolution the
eigenvalues become widely separated, that is $\rho_1\gg\cdots\gg\rho_d$, the
system is greatly simplified. In this case the antisymmetric matrix $\Omega$
becomes
\begin{eqnarray}&&
\Omega_{ik}=\left\{
\begin{array}{cc}
\tilde\sigma_{ki},\quad\quad i<k\\
-\tilde\sigma_{ik},\quad\quad i>k\,,
\end{array}
\right.
\end{eqnarray}
and due to Eqs. (\ref{prt},\ref{efd}) the dynamics of the angular degrees of
freedom is independent on the eigenvalues. Therefore, Eq. (\ref{prt}) can be
resolved:
\begin{eqnarray}&&
\rho_i=\rho_{0i}+\int_0^t \!dt'\,\tilde\sigma_{ii}(t')
\label{sol1}\\&&
+\frac{1}{2}\ln\left[
1+\kappa e^{-2\rho_{0i}}
\int_0^t\!dt'\,\exp\left\{-2\int_0^{t'}dt'\,\tilde\sigma_{ii}(t'')\right\}
\right]\,.
\nonumber
\end{eqnarray}
Here $\rho_{0i}$ are some constants of the order of unity that should be
determined by matching with the initial period of separation of the
eigenvalues.

Integrals in Eq. (\ref{sol1})  determine the dynamics of $\rho_i$. We consider
times much larger than $\tilde\tau$, the correlation time of $\tilde \sigma$,
which is generally less than or of the order of $\tau$.  The form of the
probability distribution function of $\rho$ is different if we consider
$\tilde\tau\ll t\lesssim |\lambda_d|^{-1}\ln{\rm Pe}$ or
$t\gtrsim |\lambda_d|^{-1}\ln {\rm Pe}$, depending on whether the diffusion
has started to be relevant or not. If the former case is considered, one can
disregard the second term in Eq. (\ref{sol1}), thus obtaining
$\rho_i\approx\int_0^t\!dt'\,\tilde\sigma_{ii}(t')$. We recognize the case of
the central limit theorem. However, the Gaussian distribution describes only
the bulk of the most probable events leaving rare events out of the domain of
its validity. We shall need a more general expression \cite{Frisch} which can
be derived from the following considerations. The integrals can be considered
as sums of a large number  $n\approx t/\tilde{\tau}$ of independent identically
distributed random variables. Thus we investigate the distribution of $X$
given by
\begin{eqnarray}
X=\sum_{i=1}^nx_i\,.
\label{sum}\end{eqnarray}
Without loss of the generality we can assume that
$\left\langle x_i\right\rangle=0$.
If the generating function $\left\langle\exp[iy x_i]\right\rangle$ of
each $x$ is $\exp[-s(y)]$, then $X$ has the generating function
$\exp[-ns(y)]$. To find the distribution
function of $X$, one should make the inverse Fourier transform
\begin{eqnarray}&&
P(X)=\int\frac{dy}{2\pi}\exp[iyX-ns(y)]\,.
\nonumber\end{eqnarray}
At large $n$ this integral can be calculated in the saddle-point approximation.
Writing the extremum condition, we see that $y_{\rm extr}$ is a function of
the argument $X/n$, which implies ${\cal P}(X)\propto\exp[-nS(X/n)]$. For
$X\ll n$ one can expand $S$ in the Taylor series and obtain
${\cal P}\propto\exp[-X^2/(2n \Delta)]$. Here  $\Delta$ is the variation of
$x_i$. This is nothing but the central limit theorem. On the other hand, if we
increase $n$, keeping the ratio $X/n$ a constant of the order of unity, we can
assert that $\ln{\cal P}\propto -n$. This has a simple interpretation. If $X$
is of the order of $n$, only realizations where most of $x_i$ are of the same
sign contribute. Therefore we can model the situation by the binomial random
process, which gives just the above result.

If we replace sum (\ref{sum}) by the integral $\int_0^tdt'x(t')$
of a random function $x$ over time $t$ much larger than the correlation time,
we should only note that the characteristic function of $X$ is proportional to 
$\exp[-t\tilde s(y)]$, and then proceed as above. We used the fact that the
characteristic function is an exponent of the cumulant generating function.
The derivation is easily generalized for several quantities.
Thus the distribution functions in $d=2$ and $3$ are given by the formulas
\end{multicols}
\begin{eqnarray}&&
{\cal P}\propto\exp\left[
-t\, S_2\left(\frac{\rho_1-\lambda_1 t}{t}\right)
\right]\theta(\rho_1)\delta(\rho_1+\rho_2)\,,\label{pdf2}\\&&
{\cal P}\propto\exp\left[
-t\, S_3\left(\frac{\rho_1-\lambda_1 t}{t},\frac{\rho_2-\lambda_2 t}{t}\right)
\right]\theta(\rho_1-\rho_2)\theta(\rho_2-\rho_3)\delta(\rho_1+\rho_2+\rho_3)
\,.
\label{pdf3}\end{eqnarray}
\begin{multicols}{2}
Here $S_2(x_1)$ and $S_3(x_1,x_2)$ are some functions depending on the
details of the statistics of $\sigma$. In the $\delta$ correlated case one can
find  the explicit expression for $S_{2,3}$ \cite{Son98,Kra74} (see also
Appendix \ref{Krmod}). The constants $\lambda_i$ are nothing but the Lyapunov
exponents, which are expressed via the statistics of $\sigma$ in the following
way (cf. Ref. \cite{GSO87}) 
\begin{eqnarray}&&
\lambda_i=\left\langle\tilde\sigma_{ii}\right\rangle\,.
\end{eqnarray}
To have a self-consistent picture, we should assume that the spectrum of the
Lyapunov exponents is nondegenerate, that is $\lambda_i>\lambda_{i+1}$. 
Physically nondegeneracy of the Lyapunov exponents means that a blob is
unstable with  respect to the fluctuations, leading to a separation of the
lengths of its sides. Noticing that the Lagrangian point ${\bbox r}=0$ is a
saddle point for the incompressible flow, one can easily verify that the 
strain directions corresponding to the further elongation of the blob
prevail. Therefore, during a time $\tau$ of approximately constant
strain, the blob will be on average further elongated.

At $t\gg(\lambda_i-\lambda_{i+1})^{-1}$ we can disregard effects
originating from the boundary $\rho_i=\rho_{i+1}$. Equations
(\ref{pdf2}) and (\ref{pdf3}) are not valid in a narrow region near the
boundary which has a width of the order of unity. Since it is much
smaller than $\lambda_i t$, we can use the step function $\theta$ to model the
form of the PDF near the boundary.

Due to the incompressibility condition the exponents satisfy
$\sum_{i=1}^d\!\!\lambda_i=0$. Then, in order to have a spectrum that is
nondegenerate in $d=2$, one should only require that $\lambda_1>0$, which is
the same saying that trajectories diverge exponentially. In $d=3$, it is
necessary to supply some information about $\lambda_2$. If the statistics of
$\sigma$ is symmetric with respect to time reversion, then $\lambda_2=0$
\cite{Kra74}. However, if this is not the case, it is generally nonzero. In
Appendix \ref{texp} we find the expression for $\lambda_2$ if the correlation
time of $\sigma$ is small, which shows that its sign is generally arbitrary.

The form of the functions $S_{2,3}$ depends on particular details of the
statistics of $\sigma$. However, it is possible to make two general statements
about these functions. First, one can assert that at small $x$ the
expansion
\begin{eqnarray}&&
S_2(x)\!\approx\! \frac{x^2}{2C_{11}}\,,\quad
S_3(x_1,x_2)\!\approx\! \frac{C_{22}x_1^2-2C_{12}x_1x_2+C_{11}x_2^2}
{2(C_{11}C_{22}-C_{12}^2)}
\nonumber
\end{eqnarray}
is valid, reproducing the central limit theorem.
The constants $C_{ij}$ are defined as
\begin{eqnarray}
C_{ij}=\int dt'\,\left\langle\left\langle
\tilde\sigma_{ii}(t)\tilde\sigma_{jj}(t')
\right\rangle\right\rangle\,.
\nonumber\end{eqnarray}
Here $\langle\langle\ldots\rangle\rangle$ stands for irreducible correlation
function. The integrals should be calculated over an interval,
much larger then the correlation time of $\tilde\sigma$. Note
that the condition of incompressibility ensures that $\sum_{j=1}^d C_{ij}=0$.

When $x_{i}$ are of the order of unity, the functions $S_{2,3}$ have no
singularities and change smoothly. The quadratic expansion of $S_{2,3}$ is
valid as long as 
\begin{equation}
|\rho_i-\lambda_it|\ll t/\tilde\tau\,,
\end{equation}
where $\tilde\tau$ is the correlation time of $\tilde\sigma$. In the
$\delta$-correlated case it holds everywhere (Appendix \ref{Krmod}).

The normalization of ${\cal P}$ is
determined by the quadratic part of $S_{2,3}$, since most of the probability
is concentrated at $|\rho_i-\lambda_it|\sim\sqrt{C_{ij}t}\ll t$. 
One can find the normalization factors $(2\pi C_{11}t)^{-1/2}$ in
$d=2$ and $\left[4\pi^2 t^2 (C_{11}C_{22}-C_{12}^2)\right]^{-1/2}$ in $d=3$.

Now consider $t\gtrsim |\lambda_d|^{-1}\ln{\rm Pe}$. The diffusion is
irrelevant for $\rho_i$ having a non-negative Lyapunov exponent. However, there
is a finite probability, increasing with $t$, that $\rho_d$ reaches the
diffusion scale. This requires an account of the  last term on the right-hand
side of Eq. (\ref{prt}) or Eq. (\ref{sol1}). The
diffusion will not allow $\rho_d$ to decrease much below
$\ln(|\lambda_d|/\kappa)$. On the contrary, negative $\lambda_d$ will
prevent it from increasing. As a result, the corresponding $\rho_i$ will
be distributed stationarily around the value $\ln(\kappa/|\lambda_d|)$.
Relaxation times associated with this distribution are diffusion
independent and thus are much less then $t$. On the other hand,
$\rho$ having non-negative Lyapunov exponents are the integrals over the whole
evolution time $t$, so that their values at time $t$ are not sensitive to the
last period of evolution with duration of the order of the relaxation time of
$\rho_d$. This means that fixing their values at time
$t\gg|\lambda_d|^{-1}\ln{\rm Pe}$ will not affect the distribution of
$\rho_d$, and the whole probability distribution function ${\cal P}$ is
factorized (cf. Refs. \cite{SS94,CKV97}). In $d=2$ we can write
\begin{eqnarray}&&
{\cal P}\propto\exp\left\{
-t\, S_2\left(\frac{\rho_1-\lambda_1 t}{t}\right)
\right\}{\cal P}_{\rm st}(\rho_2)\,.
\label{d=2}\end{eqnarray}
Here ${\cal P}_{\rm st}$ is the stationary distribution of $\rho_2$.
In $d=3$ the situation is more complicated. While $\lambda_3$ is always
negative, $\lambda_2$ can be both positive and negative.
The form of the PDF will be different for these two cases.
If $\lambda_2\geq0$,
\begin{eqnarray}&&
{\cal P}\propto\exp\left\{
-t\, S_3\left(\frac{\rho_1-\lambda_1 t}{t},\frac{\rho_2-\lambda_2 t}{t}\right)
\right\}{\cal P}_{\rm st}(\rho_3)\,.
\label{posit}\end{eqnarray}
Since $\rho_d$ is independent of the rest of $\rho_i$ the functions $S_{2,3}$
are the same as in Eqs. (\ref{pdf2}) and (\ref{pdf3}).

If $\lambda_2$ is negative, then at
$t\gg|\lambda_2|^{-1}\ln(|\lambda_2|/\kappa)$ the distribution over $\rho_2$
will also become steady and concentrated near $\ln(|\lambda_2|/\kappa)$ which
by order of magnitude is equal to $\ln(|\lambda_3|/\kappa)$. Therefore, our
assumption that $\rho_2\gg\rho_3$ is incorrect. Still $\rho_1\gg\rho_{2,3}$,
and the equation for $\rho_1$ is separated from the other variables.
Then the distribution function is equal to 
\begin{eqnarray}&&
{\cal P}\propto\exp\left\{
-t\, \tilde S_3\left(\frac{\rho_1-\lambda_1 t}{t}\right)
\right\}{\cal P}_{\rm st}(\rho_2,\rho_3)\,.
\label{negat}\end{eqnarray}
where $\tilde S_3$ is related to $S_3$ by
$\exp(-\tilde S_3)\propto \int d\rho_2\exp(-S_3)$.

Finally, let us note that since the configuration space ${\rm SO}(d)$ of the
rotation matrix $R$ [see Eq. (\ref{transf}] is finite, and since there is no
preferred direction in space, at large $t$ the matrix is distributed uniformly
over the sphere.

The basic result obtained above is the special scaling form
of the probability density functions. It is this universal 
form which lies in the origin of the results derived below.

\section{Decaying turbulence}\label{decay}

As a first application, let us consider decay of a passive scalar $\vartheta$.
The problem is posed as follows: given a random distribution of the scalar
density $\vartheta_0$ at $t=0$, find its statistics at $t>0$. 
In the framework of the Kraichnan model the single-point
statistics was considered by Son \cite{Son98}, who
obtained the following long-time asymptotic behavior:
\begin{eqnarray}
\left\langle|\vartheta(t,0)|^\alpha\right\rangle\propto
\exp(-\gamma_\alpha t)\,.
\label{asbe}\end{eqnarray}
where $\gamma_\alpha$ in $d=3$  is equal to $\alpha(6-\alpha)D/4$
for $0\leq\alpha<3$ and $9D/4$ otherwise ($D$ is a parameter characterizing
the strength of the fluctuations of $\sigma$). The same decay law has been
claimed for the gradients of the scalar. Here we consider the problem for
arbitrary correlation time of $\sigma$. Our consideration shows that due to
the above-mentioned special form of  Eqs. (\ref{d=2}--\ref{negat}) the law
(\ref{asbe}) is valid for an arbitrary statistics of $\sigma$ both for the
single-point value of the scalar and its gradient. In the $\delta$-correlated
limit we obtain a result for $\gamma_\alpha$ different from that of Ref.
\cite{Son98}. The results also show that the basic assumption of Ref.
\cite{SY} is incorrect.  

The following qualitative picture, supported by the calculations presented
below, explains the decay. First, consider a single blob initially having a
characteristic size $L$ and containing $N$ particles of the scalar. As
velocity stretches the blob, the number of particles does not change, contrary
to the volume of the blob. At
$t\gtrsim |\lambda_d|^{-1}\ln(|\lambda_d|L^2/\kappa)$ the dimensions of the
blob with negative Lyapunov exponents are frozen at $r_{\rm dif}$, while the
rest keep growing exponentially, resulting in an exponential growth of the
total volume of the blob. Since the volume is proportional to $\sqrt{\det I}=
\exp\left(\sum\rho_i\right)$, one has $\left\langle|\vartheta|^{\alpha}
\right\rangle \propto\left\langle\exp\left(-\alpha\sum\rho_i\right)
\right\rangle$, where
the averaging should be done with the help of the PDF discussed above,
that is $\left\langle|\vartheta|^{\alpha}\right\rangle\propto
\int d\rho \,{\cal P}(\rho)\exp\left(-\alpha\sum\rho_i\right)$.
The result is determined by a compromise between two competing factors.
While the averaged quantity $\exp\left(-\alpha\sum\rho_i\right)$ favors
smaller values of $\sum\rho_i$, the maximum of the probability is attained
when each growing $\rho_i$ is equal to $\lambda_i t$. Obviously,
for larger $\alpha$ the volume acquires more importance, so that the main
contribution is made by smaller blobs, which are less probable but have
larger concentration of the scalar. 
So, at small $\alpha$, the deviation from
the average growth $\lambda t$ is small, and $\gamma_\alpha$ is determined by
the Gaussian part of the PDF. This gives a parabolic dependence on $\alpha$.
On the other hand, if $\alpha$ is large enough, the main contribution is due
to a blob having a minimal possible volume which is of the order of $L^d$.
The decay exponent is fully determined by the probability to have such
a blob and hence is $\alpha$ independent \cite{SS94,Son98}. Note that this
picture implies that the exponential decay holds at
$t\gtrsim|\lambda_d|^{-1}\ln{\rm Pe}$.

If instead of a single blob one takes a spatially homogeneous problem, this
consideration should be slightly modified. At large $t$, initially uncorrelated
blobs are brought close to each other because of the contraction along a
certain direction, and then they overlap diffusively. Since the number of
overlapping blobs is large, due to the central limit theorem it is rather
$\vartheta^2$ which is inversely proportional to volume. Therefore, 
$\left\langle|\vartheta|^\alpha\right\rangle\propto\left\langle
\exp\left(-\alpha\sum\rho_i/2\right)\right\rangle$.

Formally, one should solve Eq. (\ref{pseq}) with the initial condition
$\vartheta(0,{\bbox r})=\vartheta_0({\bbox r})$. The solution is
\begin{eqnarray}&&
\vartheta(t,0)\!=\!\int\frac{d{\bbox k}}{(2\pi)^d}
\vartheta_0\left(W^{T}(t){\bbox k}\right)
\exp\left[-Q_{\mu\nu}k_\mu k_\nu\right]\,,
\label{theta_0}\\&&
Q(t)=\kappa\int_0^t\!dt'\, W(t)W^{-1}(t')\left[W(t)W^{-1}(t')\right]^{T}\,.
\end{eqnarray}
From the qualitative arguments it is clear that the long-time asymptotic
should be independent of the particular form of the distribution. We will
take the simplest statistics, which is Gaussian with the pair correlation
function
\begin{eqnarray}&&
\left\langle\vartheta_0({\bbox r}_1)\vartheta_0({\bbox r}_2)\right\rangle\!=\!
\chi(r_{12})\,,\quad
\chi\!=\!\chi_0 \exp[-r^2/(8L^2)]\,.
\label{chi}\end{eqnarray}
This particular form is chosen for further convenience. In what follows we set
$L=1$. It is possible to generalize the calculation for arbitrary $\chi$ and
show that the results are independent of its form. 

To proceed, we introduce the generating function of $\vartheta$:
\begin{eqnarray}&&
{\cal Z}(y)=
\left\langle\exp[iy\vartheta(t,0)]\right\rangle_{\sigma,\vartheta_0}\,.
\label{Z1}\end{eqnarray}
Here we assume averaging over the statistics of $\sigma$ and the initial
distribution of the scalar. The simplest part is to perform averaging over the
Gaussian field $\vartheta_0$. To do this, we substitute Eq. (\ref{theta_0})
into Eq. (\ref{Z1}) and using the expression for the characteristic function
of a Gaussian random variable \cite{Frisch}, obtain
\begin{eqnarray}&&
{\cal Z}(y)=
\left\langle
\exp\left[-\frac{y^2}{2}
\int\frac{d{\bbox k}}{(2\pi)^d}\chi(W^T{\bbox k})
e^{-2Q_{\mu\nu}k_\mu k_\nu}
\right]\right\rangle_{\sigma}\,.
\nonumber\end{eqnarray}
Substituting $\chi(k)=(8\pi)^{d/2}\chi_0\exp(-2k^2)$ and integrating
over ${\bbox k}$ we obtain
\begin{eqnarray}&&
{\cal Z}=\left\langle\exp\left[-\frac{y^2}{2}\frac{\chi_0}
{\sqrt{\det I(t)}}
\right]\right\rangle_{\sigma}\,.
\label{Z}
\end{eqnarray}
We used the fact that $I=WW^T+Q$, which can be verified by writing equation
on $WW^T+Q$ and comparing it with Eq. (\ref{I2}).

Using Eq. (\ref{Z}), one can find
\begin{eqnarray}&&
\left\langle|\vartheta(t,0)|^\alpha\right\rangle=
C_\alpha\left\langle
(\det I)^{-\alpha/4}\right\rangle_\sigma\,.
\label{U}\end{eqnarray}
Here $C_\alpha$ is a numerical constant.
Equation (\ref{U}) reduces the problem to an averaging of powers of
$\det I=\prod\exp(2\rho_i)$,
\begin{eqnarray}&&
\left\langle|\vartheta|^\alpha\right\rangle=C_\alpha
\int d^d\rho\,\,\exp\left[-\frac{\alpha}{2}\sum_{i=1}^d\rho_i\right]
{\cal P}(t,\rho)\,,
\label{wcw}\end{eqnarray}
where ${\cal P}$ is the probability density function of $\rho$ discussed
above. In the large time limit this integral can be calculated in the
saddle-point approximation. The calculation is slightly different for $d=2$
and $3$.

\subsection{Two-dimensional case}

In $d=2$ integral (\ref{wcw}) should be calculated only over $\rho_1$,
since the distribution over $\rho_2$ is stationary.
The saddle-point equation is
$$
S_2'\left(\frac{\rho_1-\lambda_1 t}{t}\right)+\frac{\alpha}{2}=0\,.
$$
It is clear that $\rho_1\propto t$.
As long as one can use the quadratic expansion, which valid at least at small
$\alpha$, the solution of this equation is
$\rho_{1}=(\lambda_1-\alpha C_{11}/2)t$, hence
\begin{eqnarray}&& 
\gamma_\alpha=\frac{\alpha}{2}\left(\lambda_1-\frac{\alpha C_{11}}{4}\right)
\,.
\label{fras}\end{eqnarray}
At $\alpha>\alpha_{\rm cr}=-2S_2'(-\lambda_1)$
the value of $\rho_{1}$ becomes
much smaller than $\lambda_1t$; the integral (\ref{wcw}) is determined by
the boundary of the integration region, and therefore
$\gamma_\alpha=S_2(-\lambda_1)$, independent of $\alpha$.

The domain of validity of Eq. (\ref{fras}) depends on the value of the
parameter $\lambda\tilde\tau$. If it is much smaller than unity, we can use
the quadratic approximation to $S_2$ everywhere. This case effectively
corresponds to the Kraichnan limit \cite{Son98}.

In the opposite limit $\lambda\tilde\tau\gtrsim 1$, the quadratic
expansion of $S_2$ cannot be used for $\alpha\gtrsim 1/(\tilde\tau C_{11})$,
and Eq. (\ref{fras}) is valid only at $\alpha\ll 1/(\tilde\tau C_{11})$.
The form of the intermediate region is not universal, and depends on the
particular form of $S_2$ and hence on details of the statistics of $\sigma$.

\subsection{Three-dimensional case}\label{three}

In $d=3$ the result is similar to that of $d=2$, though the consideration
is slightly more complicated, due to the presence of an additional degree of
freedom. There are two cases to be considered. If $\lambda_2<0$, then 
$\cal P$ is given by Eq. (\ref{negat}) and the calculation is the same
as for $d=2$. 

If $\lambda_2\geq0$, both degrees of freedom $\rho_1$ and $\rho_2$ are active
and one should use the PDF (\ref{posit}). The saddle-point equations are
\begin{eqnarray}&&
\frac{\partial S_3(x_1,x_2)}{\partial x_1}+\frac{\alpha}{2}=0\,,\quad
\frac{\partial S_3(x_1,x_2)}{\partial x_2}+\frac{\alpha}{2}=0\,,
\label{sp1}\end{eqnarray}
where $x_1=(\rho_1-\lambda_1 t)/t$ and $x_2=(\rho_2-\lambda_2 t)/t$.
Again, the beginning of the curve is determined by the Gaussian
part of $\cal P$, and $\gamma_\alpha$ is parabolic:
\begin{eqnarray}&&
\gamma_\alpha=\frac{\alpha}{2}\left(|\lambda_3|-\frac{\alpha}{4}
C_{33}\right)\,,
\label{Gau}\end{eqnarray}
with $\rho_{1}=(\lambda_1+\alpha C_{13}/2)t$ and
$\rho_2=(\lambda_2+\alpha C_{23}/2)t$.
At $\alpha$ larger than $\alpha_{\rm cr}$ calculated below, we have the
$\alpha$-independent behavior
\begin{equation}
\gamma_\alpha=S_3(-\lambda_1,-\lambda_2)\,.
\end{equation}

Depending on the parameters, two different types of behavior can occur at
$\alpha<\alpha_{\rm cr}$. First, it is possible that as $\alpha$ increases,
$\rho_2$ will grow more slowly with $t$  and at certain $\alpha$ will become
much smaller than $\lambda_2 t$. At larger $\alpha$ the integration over
$\rho_2$ will be determined by the region $\rho_2\ll \lambda_2 t$ and one
should replace system (\ref{sp1}) by the single equation
\begin{eqnarray}&&
\frac{\partial S_3(x_1,-\lambda_2)}{\partial x_1}+\frac{\alpha}{2}=0\,.
\end{eqnarray}
In this case $\alpha_{\rm cr}=-2\left.\partial_{1}
S_3(x_1,-\lambda_2)\right|_{x_1=-\lambda_1}$.

The other possibility is that at certain $\alpha$ the difference
$\rho_1-\rho_2$ becomes much smaller than $\rho_{1,2}$. 
Because of the constraint $\rho_1>\rho_2$, for larger $\alpha$
integral (\ref{wcw}) is determined by the boundary
$\rho_1=\rho_2$ of the domain of integration and the saddle-point equation
becomes
\begin{eqnarray}&&
\frac{\partial S(x_1,x_2)}{\partial x_1}+
\frac{\partial S(x_1,x_2)}{\partial x_2}+\alpha=0\,.
\end{eqnarray}
Then
$\alpha_{\rm cr}=-\left[\partial_1 S(x_1,x_2)+\partial_2 S(x_1,x_2)\right]$
at $x_1=-\lambda_1$ and $x_2=-\lambda_2$. Geometrically, the first case
corresponds to elongated ellipsoids, and the second one to sheets having two
largest dimensions of the same order.

If the parameter $\lambda \tilde\tau$ is small enough, these changes of the
regime occur within the Gaussian part of $S_3$. Then one can perform a more
detailed investigation. The first
regime is realized if $\lambda_1>\lambda_2C_{13}/C_{23}$ and $C_{23}<0$. Then,
at $\alpha>-2\lambda_2/C_{23}$  Eq. (\ref{Gau}) should be replaced by
\begin{eqnarray}&&
\gamma_\alpha\!=\!
\frac{\alpha^2}{8}\left(\frac{C_{12}^2}{C_{22}}\!-\!C_{11}\right)
\!+\!\frac{\alpha}{2}\left(
\lambda_1-\frac{C_{12}\lambda_2}{C_{22}}
\right)\!+\!\frac{\lambda_2^2}{2C_{22}}\,.
\label{Gau2}\end{eqnarray}
If $\lambda_2=0$, Eq. (\ref{Gau}) has no region of validity,
and at $\alpha>0$ one should use Eq. (\ref{Gau2}) which becomes 
\begin{eqnarray}&&
\gamma_\alpha=\frac{\alpha}{2}\left[\lambda_1-
\frac{\alpha}{4}\left(C_{11}-\frac{C_{12}^2}{C_{22}}\right)\right]\,.
\label{Gau3}\end{eqnarray}
In particular, within the Kraichnan model, Eq. (\ref{Gau3}) is valid
for $0\leq\alpha\leq\alpha_{\rm cr}$. Substituting $\lambda_1$ and $C_{ij}$
(see Appendix \ref{Krmod}) one finds $\alpha_{\rm cr}=4$,
\begin{eqnarray}&&
\gamma_{\alpha}=\frac{3D\alpha}{2}\left(1-\frac{\alpha}{8}\right)\,,
\label{Son}\end{eqnarray}
for $\alpha<\alpha_{\rm cr}$ and $\gamma_{\alpha}=3D$ for
$\alpha>\alpha_{\rm cr}$. Our result is different from the one obtained in
Ref. \cite{Son98}, which coincides rather with Eq. (\ref{Gau}). An exact
solution for $\alpha=2$ (see Appendix \ref{Pair} and Ref. \cite{Fou98})
supports 
Eq. (\ref{Son}). The reason for the discrepancy is the following. Despite the
fact that $\rho_2\propto\ln(\kappa/D)$, it is impossible to ignore it 
completely. If this were done, the anticorrelation between $\rho_1$ and
$\rho_2$, existing due to the incompressibility condition would lead to the
growth of $\rho_2$, thus making the calculation inconsistent.

The second regime takes place if $C_{23}>C_{13}$ and
$C_{23}>C_{13}\lambda_2/\lambda_1$. Then, starting from
$\alpha=2(\lambda_1-\lambda_2)/(C_{23}-C_{13})$ Eq. (\ref{Gau}) should
be replaced by another formula. Although the dependence on $\alpha$ 
is still parabolic, the coefficients are rather cumbersome, so we do not
write this here.

\subsection{Gradients of the decaying scalar}

In the same manner one can consider the decay of the gradients of the scalar.
In analogy, we can look for the correlation functions
$\left\langle |{\bbox\omega}|^\alpha\right \rangle$, where
${\bbox \omega}=\left.\nabla\vartheta(t,{\bbox r})\right|_{r=0}$. As in the
case of single-point scalar statistics, these correlation functions decay
exponentially in time. It was claimed in Ref. \cite{Son98} that the decay law
of the scalar and its gradient is the same within the Kraichnan model. Here we
show that this is actually the case for arbitrary correlated strain.
Qualitatively it follows from the estimate that
$|\nabla \vartheta|\approx \vartheta/l$, where $l=\exp(\rho_d)$ is the smallest
dimension of  the blob. As explained above, $\vartheta$ and $l$ can be
considered as independent,  while the statistics of $l$ is stationary. Thus the
decay of the gradient is solely due to the change of the density of the scalar.
More formally, one has
\begin{eqnarray}&&
\omega_\alpha
=i\int\frac{d{\bbox k}}{(2\pi)^d}k_\alpha
\vartheta_0\left(W^{T}(t){\bbox k}\right)
\exp\left[- Q_{\mu\nu}k_\mu k_\nu\right]
\end{eqnarray}
Introducing the  function ${\cal Z}({\bbox y})= \left\langle
\exp\left[i\left({\bbox y},{\bbox \omega}\right)\right]\right\rangle$, and
averaging it over the initial distribution (\ref{chi}) we obtain a
formula similar to Eq. (\ref{Z}) Then, making a Fourier transform over
${\bbox y}$, we obtain the PDF of ${\bbox \omega}$:
\begin{eqnarray}&&
{\cal P}\propto
\left\langle
(\det I)^{d/4+1/2}\exp\left[-\frac{\sqrt{\det I}}
{\chi_0}\,({\bbox \omega},
I{\bbox \omega})\right]
\right\rangle_\sigma\,.
\label{Zw}\end{eqnarray}
Considering this expression in the eigenbasis of the matrix $I$, we observe
that $\langle|{\bbox \omega}|^\alpha\rangle\sim
\langle|\omega_d|^\alpha\rangle$, since $\rho_d$ is smaller than the rest of
the $\rho$ is. Recalling that the distribution over $\rho_d$ is stationary, we
immediately obtain that
\begin{eqnarray}&&
\left\langle |\nabla_\alpha\vartheta(t,0)|^\alpha\right\rangle
\propto\left\langle(\det I)^{-\alpha/4}\right\rangle_\sigma\,,
\nonumber\end{eqnarray}
which, due to Eq. (\ref{U}), gives the same law of decay.

\section{Forced turbulence}\label{forced}
\subsection{Single-point distribution of $\vartheta$}\label{sp}

In this section we shall investigate the steady state distribution of a passive
scalar which occurs in the presence of a stationary source.
For this purpose we introduce a random function $\phi(t,{\bbox r})$
on the right-hand side of  Eq. (\ref{pseq}),
injecting blobs of the scalar with the characteristic size $L$.
Due to the linearity of the problem, the scalar field at the moment $t=0$
is given by a superposition of the scalar injected at earlier instants of time.
Each realization of $\sigma$ can be characterized by a parameter $t_*$
(cf. Ref. \cite{SS94}), such that the smallest dimension of blobs injected at
$t\approx-t_*$ approaches $r_{\rm dif}$ at $t=0$.  The ambiguity in the
definition of $t_*$ is of the order of $|\lambda_d|^{-1}$
which is much smaller than the typical stretching time
$|\lambda_d|^{-1}\ln{\rm Pe}$. 
Considering the motion of the scalar injected at $-t_*\lesssim t<0$ one may
neglect diffusion, and the scalar is simply transfered along Lagrangian
trajectories.  On the other hand, as discussed in Sec. \ref{decay}, the
contribution of the scalar injected at  $t\lesssim -t_*$ is exponentially
small. Thus $t_*$ separates diffusive and diffusionless regimes.
One can write the following approximate formula:
\begin{eqnarray}&&
\left. \vartheta(0,{\bbox r})\right|_{r=0}
\approx\int_{-t_*}^0\!dt\,\phi(t,0)\,. 
\label{appr}\end{eqnarray}
If the correlation time of  the source is much smaller than $t_*$,
for a fixed realization of $\sigma$ integral (\ref{appr}) can be considered
as a Gaussian variable with zero average and the dispersion proportional
to $t_*$, so that after averaging over $\phi$,  for
the single-point PDF \cite{SS94} one obtains
\begin{eqnarray}&&
{\cal P}(\vartheta)=\left\langle \frac{1}{\sqrt{2\pi \chi_0 t_*}}
\exp\left(-\frac{\vartheta^2}{2\chi_0 t_*}\right)
\right\rangle_\sigma
\label{shra}\end{eqnarray}
where $\chi_0=\int dt\left\langle\phi(t,0)\phi(0,0)\right\rangle$. The
effective Gaussianity of the pumping has its limitations due to finite
correlation time of $\phi$ [analogously to the discussion of Eqs.
(\ref{pdf2}) and (\ref{pdf3})]. Since we work in the comoving reference frame,
this correlation time is very small, and hence only the tail will be affected.
At the end of the section we discuss the implications of this. To proceed with
formal calculations, we should specify statistics of $\phi$. Here we shall
take $\phi$ as a Gaussian field with the pair-correlation function
$$
\left\langle\phi(t_1,{\bbox r}_1)\phi(t_2,{\bbox r}_2)\right\rangle=
\chi(r_{12})\delta(t_1-t_2)\,,
$$
where $\chi(r)$ is the same as in sec. \ref{decay}.
The expression for the generating function
${\cal Z}=\left\langle\exp(iy\vartheta)\right\rangle$ follows
\begin{eqnarray}&&
{\cal Z}=\left\langle
\exp\left[
-\frac{y^2\chi_0}{2}\int_{-\infty}^0 \! \frac{dt'}{\sqrt{\det{I(0,t')}}}
\right]\right\rangle_{\sigma}\,.
\label{Z2}\end{eqnarray}
Here $I(t,t')$ is a matrix, satisfying Eq. (\ref{I2}) with respect to $t$
and the initial condition $I(t',t')=1$. Naturally, the integration is
performed over the initial time, summing up the blobs injected at different
times.
The integral in Eq. (\ref{Z2}) gives the formal definition of $t_*$
entering Eq. (\ref{shra}):
\begin{equation}
t_*=\int_{-\infty}^0 \!\frac{dt'}{\sqrt{\det I(0,t')}}\,.
\end{equation}
Introducing the distribution function $p(t_*)$, we rewrite Eq. (\ref{shra}) as
\begin{eqnarray}&&
{\cal P}(\vartheta)=\int_0^\infty \!\frac{dt_*}{\sqrt{2\pi\chi_0 t_*}}
\,p(t_*)\exp\left(
-\frac{\vartheta^2}{2\chi_0t_*}
\right)\,.
\end{eqnarray}

Since $t_*$ is a functional of the whole trajectory $\rho_i(t,t')$,
one needs more information than contained in the simultaneous
distribution function of $\rho$.
However, the following approximation, becoming exact at $\ln{\rm Pe}\to\infty$,
reduces the problem to single-time statistics. We shall neglect the
configurations for which the smallest dimension of the blob once reached
$r_{\rm dif}$ starts to grow. Then the realizations for which $t_*$ is larger
than some $T$, and those for which the blob injected at $-T$ has
$\rho_3(0,-T)>r_{\rm dif}$, are the same, leading us to the following formulas:
\begin{eqnarray}&&
p(t_*)=\frac{\partial}{\partial t_*}
\int_{-\infty}^\infty d\rho_1\int_{\ln(\kappa/|\lambda_2|)}^\infty
\!\!d\rho_2\,\,
{\cal P}(t_*,\rho_1,\rho_2)\,,
\label{esc2}
\\&&
p(t_*)\!=\!\frac{\partial}{\partial t_*}\int_{-\infty}^\infty \!\!\!d\rho_1\!
\int_{-\infty}^\infty \!\!\!d\rho_2\!
\int^{\infty}_{\ln(\kappa/|\lambda_3|)} \!\!\!\! d\rho_3
\,\,{\cal P}(t_*,\rho_1,\rho_2,\rho_3)\,.
\nonumber\end{eqnarray}
Here one should substitute PDF's (\ref{pdf2}) and (\ref{pdf3}), since $t_*$ is
determined by the diffusionless regime. These equations define nothing but
the flux of the probability out of the region
$\rho_3>\ln(\kappa/|\lambda_d|)$, once returns are disregarded. 

Investigation of the above integrals shows that $p(t_*)$ has the following
properties. Its main body is concentrated in the vicinity of
$t_*=|\lambda_d|^{-1}\ln{\rm Pe}$, and has a width
of the order of $\sqrt{\ln{\rm Pe}}$. On the other
hand, its tail $t_*\gg|\lambda_d|^{-1}\ln{\rm Pe}$ decays
exponentially,
\begin{eqnarray}&&
p(t_*)\propto \exp(-ct_*)\,,
\label{exp}\end{eqnarray}
where $c$ is equal to
$S_2(-\lambda_1)$ in $d=2$ and $S_3(-\lambda_1,-\lambda_3)$ in $d=3$. The
intermediate region is not universal and depends on the details of $\sigma$.
Note that since Eq. (\ref{exp}) gives the probability that at large
time $t_*$ a blob has not yet decayed, therefore $c$ is equal to
the limiting value of $\gamma_\alpha$ (see sec. \ref{decay}).

This information allows one to calculate the probability distribution function
${\cal P}(\vartheta)$. If $\vartheta\ll\ln(|\lambda_d/\kappa)$,
it is the central peak of $p(t_*)$ that determines the scalar PDF:
\begin{eqnarray}&&
{\cal P}\!=\!
\left(\frac{|\lambda_d|}{2\pi\chi_0\ln{\rm Pe}}\right)^{1/2}
\exp\left[-\frac{|\lambda_d|\vartheta^2}
{2\chi_0\ln{\rm Pe}}\right]\,.
\label{gau2}\end{eqnarray}
As we increase $\vartheta$, at
$\vartheta\gtrsim\ln{\rm Pe}$ the details of the distribution of
$t_*$ become
important. The Gaussian regime [Eq. (\ref{gau2})] will turn into some
nonuniversal asymptotic. Nonetheless, at $\vartheta\gg \ln{\rm Pe}$,
due to Eq. (\ref{exp}) the universality is restored:
\begin{eqnarray}&&
{\cal P}\propto\exp\left(-\sqrt{\frac{2c}{\chi_0}}\,|\vartheta|\right)\,.
\label{Expon}
\end{eqnarray}

For a $\delta$-correlated $\sigma$ one can find the complete function
${\cal P}(\vartheta)$. Then $S_{2,3}$ is Gaussian (Appendix \ref{Krmod})
and the result can be found in the saddle- point approximation.
In $d=2$ the result coincides with that of Refs. \cite{CFKL95,KLS98,CGK94}. In
$d=3$ we obtain the formula (cf. Refs. \cite{KLS98,CGK94})
\begin{eqnarray}&&
\ln{\cal P}\propto\
-3\left[\sqrt{\ln^2{\rm Pe}+
\frac{D\vartheta^2}{2\chi_0}}-
\ln{\rm Pe}\right]
\nonumber\end{eqnarray} 
for $|\vartheta|<4\sqrt{\chi_0/D}\ln{\rm Pe}$, and
\begin{eqnarray}&&
\ln{\cal P}\propto 6\ln {\rm Pe}-4\sqrt{ 3\ln^2{\rm Pe}+
\frac{3D\vartheta^2}{8\chi_0}}
\nonumber\end{eqnarray}
otherwise. The change of the regime is related to the fact that the
two dimensions of the contributing blobs start to be equal to $r_{\rm dif}$.
This result is different from the one presented in Ref. \cite{KLS98}.
The difference can be qualitatively explained as follows.
In our case the structures of the scalar making the main contribution to
the PDF are columns, with the two smallest dimensions of the same
order. They appear because of the anticorrelation originating from
the incompressibility condition: for $t_*$ larger than a mean value, $\rho_3$
should decrease slower than $\lambda_3t$, which by virtue of the
anticorrelation leads to a decrease of $\rho_2$ faster than $\lambda_2t$.
Staring from a certain value of $\vartheta$, both $\rho_2$ and $\rho_3$ 
decrease at the same rate (an analogous phenomenon is described in subsection
\ref{three}). This structure is different from the ansatz proposed in
Ref. \cite{KLS98}.

Equation (\ref{shra}) should be modified if $\phi$ has a finite correlation
time $\tau_\phi$ [see the discussion leading to Eqs. (\ref{pdf2}] and
(\ref{pdf3}) ). That is
$$
{\cal P}=\left\langle\frac{1}{\sqrt{2\pi\chi_0 t_*}}\exp\left[-t_*f\left(
\frac{\vartheta}{t_*}\right)\right]
\right\rangle_\sigma,
$$
where $f(x)$ deviates from $x^2/(2\chi_0)$ at $x\gtrsim 1/\tau_\phi$.
This may affect only the tail of ${\cal P}(\vartheta)$. 
If the parameter $\tau_\phi\sqrt{\chi_0 c}$ is much smaller than unity
the tail is determined completely by the region where $f\propto x^2$,
and one obtains the asymptotic result (\ref{Expon}).
Conversely, at $\tau_\phi\sqrt{\chi_0 c}\gtrsim 1$ the form
of $f$ should be accounted. Nevertheless, one can easily check that the
exponential tail survives with a decrement depending on the form of $f$.

\subsection{Gradients}

Here we briefly consider the statistics of the scalar gradients. Within
the Kraichnan model the problem was solved in Ref. \cite{CFK98}. From the
qualitative picture presented there, one can conclude that
the PDF is determined by the short-time fluctuations of $\sigma$, and hence
is nonuniversal. The following considerations support the conclusion.

In a way, similar to the one leading to Eqs. (\ref{Z}), (\ref{Zw},
(\ref{shra}, and (\ref{Z2}) one can find
\begin{eqnarray}&&
{\cal Z}({\bbox y})=\left\langle
\exp\left[
-\frac{y_\alpha y_\beta\chi_0}{4}\int_{-\infty}^0
\frac{I_{\alpha\beta}^{-1}(0,t')\,dt'}{\sqrt{\det I(0,t')}}
\right]
\right\rangle_\sigma\,.
\label{Z5}
\end{eqnarray}
Analogously to the case of the scalar density, the gradient field at $t=0$
is given by a superposition of contributions of blobs injected at earlier
moments of time. We observe that the contribution of each blob into
$\omega^2\equiv({\bbox \nabla}\vartheta)^2$ is determined by two factors: the
value  of the scalar density $\chi_0(\det I)^{-1/2}$ and the inverse size of
the blob contained in $I_{\alpha\beta}^{-1}$. Not all the blobs make a
contribution to Eq. (\ref{Z5}). Indeed, the size of the blobs injected at
$|t'|\ll |t_*|$, (where $t_*$ is defined in subsection \ref{sp}) is much
larger than $r_{\rm dif}$ and, therefore the value of the gradient will be
small. On the other hand, the scalar injected at $|t'|\gg|t_*|$ has an
exponentially small density and hence does not contribute. Thus the
distribution of $\omega$ is determined by the blobs injected at
$t'\approx -t_*$ which have the minimum possible size
provided the diffusion is still ineffective at $t=0$.

Each realization of $\sigma$ can thus be roughly characterized by two relevant
parameters. The first one is the lateral dimension $l$ of the thinnest blobs,
for which the diffusion can still be neglected at $t<0$. This is related to the
very last stage of the evolution, when blobs of the smallest size of the
order of $r_{\rm dif}$ may undergo a strong rapid contraction, increasing the
gradient without dissipating the scalar. Let us stress that the fluctuation
should be short lived in order to suppress the diffusive spreading of
the particles.

The other parameter is the duration $T_0$ of the injection stage for these
blobs, showing how many blobs approach $r_{\rm dif}$ at $t\approx0$.
There is no average strain during this period, so that blobs injected at
$-t_*-T_0\lesssim t'\lesssim -t_*$ all have a size of the order $L$ at
$t\approx -t_*$. Since at $t>-t_*$ the blobs move in the same velocity
field, they all have approximately the same size at $t=0$.
Formally, the number of relevant blobs is expressed by the formula
[c.f. Eq. (\ref{appr})]
\begin{equation}
\vartheta\approx \int_{-T_0-t_*}^{-t_*} \!\!\!dt'\,\phi(t',0)\,.
\label{inj}\end{equation}
Writing the estimation for the gradient $\omega\approx\vartheta/l$
we can replace Eq. (\ref{Z5}) by
${\cal Z}=\left\langle\exp(iy\vartheta/l)\right\rangle_{\phi,\sigma}$.
Averaging over $\phi$, we obtain
$$
{\cal Z}=\left\langle\exp\left(-\frac{y^2\chi_0T_0}{2l^2}\right)
\right\rangle_{T_0,l}\,.
$$
Since the injection stage occurs at $|t|\gtrsim |t_*|\approx
|\lambda_d|^{-1}\ln{\rm Pe}\gg\tilde\tau$, the  fluctuations determining
$T_0$ and $l$ are independent \cite{CFK98,CKV97}. 

On the average $T_0\sim |\lambda_d|^{-1}$
and $l\sim r_{\rm dif}$, so that $\left\langle\omega^2\right\rangle
\sim \chi_0/(\lambda_d^{-1}r_{\rm dif}^2)$.
Nevertheless, studying the tail of the gradients PDF, it is necessary
to take into account the large deviations of these parameters. The probability
of a large value of $T_0$ is related to configurations of
small strain (see Secs. \ref{decay} and \ref{sp}), and decays as
$\exp(-cT_0)$. Writing
$$
{\cal P}(\omega)\sim\left\langle\exp[-\omega^2 l^2/(2\chi_0 T_0)]
\right\rangle_{T_0,l}\,,
$$
one can average over $T_0$ and find 
$$
{\cal P}(\omega)\sim\left\langle\exp[-|\omega|l (2c/\chi_0)^{1/2}]
\right\rangle_l\,.
$$ 
To average over $l$ one  notes that the tail $l\ll r_{\rm dif}$ of the
probability distribution
function of $l$ is related to the tail of ${\cal P}_{\rm st}(\rho_d)$
[see Eqs. (\ref{d=2}--\ref{negat})] via $l=\exp(\rho_d)$. Indeed, both are
determined by the probability of a strong and rapid contraction
from $r_{\rm dif}$ to $l\ll r_{\rm dif}$. Hence
\begin{equation}
{\cal P}(\omega)\sim \int\! dl\,{\cal P}_{\rm st}(\ln l)
\exp\left[-|\omega| l(2c/\chi_0)^{1/2}\right]
\end{equation}
Within the Kraichnan model $\ln{\cal P}_{\rm st}\propto -l^{-2}$ (see Appendix A)
and the result $\ln {\cal P}(\omega)\propto -|\omega|^{2/3}$ of Ref.
\cite{CFK98} easily follows. In general the fluctuations of the smallest
dimension take place at times of order $\tau$ near $t=0$, and therefore are
related to the single-time distribution of $\tilde{\sigma}$, which is
nonuniversal.

We conclude that the gradient statistics is nonuniversal, and cannot be
predicted unless some specific information is supplied.   
For example, if the distribution of $l$ falls off very fast at small $l$, 
the distribution of $\omega$ will have an exponential tail at very large
$\omega$. In particular, this can explain the results of numerical
simulations \cite{SS94,BDFun}, where a cutoff can be due either to the
grid step or imposed by hand \cite{SS94}.
If the tail of ${\cal P}_{\rm st}$ behaves according to
$\ln[{\cal P}_{\rm st}(l)]\propto -l^{-\alpha}$, the tail
of ${\cal P}(\omega)$ has the stretched exponential form:
$\ln[{\cal P}(\omega)]\propto -|\omega|^{\alpha/(\alpha+1)}$.

\section{Conclusion}

We considered a passive scalar advected by a random
large-scale velocity field.  Our purpose was to establish the degree of
universality of the scalar statistics for arbitrary an correlated velocity.
The investigation can be reduced to the statistics of different Lagrangian
characteristics of the smooth flow. In the limit of large a Peclet number, part
of the relevant information is contained in the long-time asymptotic
properties of the Lagrangian statistics, which is shown to possess a
universal form. The scalar quantities related to long-time
evolution thus manifest universal statistical features. Generally, these are
the central part and the tail of the corresponding PDF.
We considered several particular examples of such quantities:
the decay of the scalar density and its gradient, and the
scalar density in the forced case. Conversely, the statistics of the
gradients in the forced case requires information about short-time
fluctuations of the velocity, and is thus sensitive to its details. 

The application of the Lagrangian statistics established here is not only
restricted to the above examples. One can slightly modify the procedure to
consider many-point correlation functions of the scalar, say the PDF of
the scalar difference at two points \cite{KLS98}, correlation functions
out of the convective interval \cite{BFLL99}, and other problems.
A similar scheme could be applied to other passive
quantities, like vectors \cite{CFKV} and tensors \cite{Lum72}.

\acknowledgments

We wish to express our gratitude to G. Falkovich and 
V. Lebedev for numerous helpful discussions and comments. Remarks
by I. Kolokolov, M. Vergassola, Y. Levinson and J. L. Gilson are
greatly acknowledged.

\end{multicols}

\appendix

\begin{multicols}{2}

\section{The Kraichnan model}\label{Krmod}

Here we consider Eqs. (\ref{prt},\ref{efd}) within the Kraichnan
model, when matrix $\sigma$ has zero correlation time and is Gaussian with the
pair-correlation function \cite{Kra68}
\begin{eqnarray}&&
\left\langle\sigma_{\alpha\beta}(t)\sigma_{\mu\nu}(0)\right\rangle
\!=\!D[(d+1)\delta_{\alpha\mu}\delta_{\beta\nu}
\!-\!\delta_{\alpha\beta}\delta_{\mu\nu}\!-\!\delta_{\alpha\nu}
\delta_{\beta\mu}
]\delta(t)\,.
\nonumber\end{eqnarray}
The tensor structure is fixed by incompressibility condition.
Zero correlation time allows to write the Fokker-Planck equation for the
probability distribution of $R$ and $\Lambda$. Integrating out the angular
degrees of freedom, one can see that the equation obtained is
equivalent to the following Langevin dynamics \cite{BGK97}
\begin{eqnarray}&&
\partial_t\rho_i=\frac{Dd}{2}\sum_{j\not=i}^d
\coth(2\rho_i-2\rho_j)+\xi_i+\frac{\kappa}{2}\exp(-2\rho_i) \,,
\nonumber
\end{eqnarray}
where $\xi_i$ are random Gaussian delta-correlated processes
with the following correlation functions
\begin{eqnarray}&&
\left\langle\xi_i(t)\xi_j(t')\right\rangle=C_{ij}\delta(t-t')\,,\quad
C_{ij}=D(d\delta_{ij}-1)\,.
\nonumber
\end{eqnarray}
Let us now consider a typical evolution of the eigenvalues.
At $t=0$ all the eigenvalues $\rho_i$ are equal to zero. Then, 
during short initial period of time, all $\rho$ start to differ. We can always
arrange $\rho$ so that $\rho_1>\rho_2>..>\rho_d$. We observe that then the
ballistic terms $\sum\coth(2\rho_i-2\rho_j)$ are arranged in the same order so
that the eigenvalues will continue to separate and at $t\gg D^{-1}$ the
following inequalities will hold $\rho_1\gg \rho_2\gg\ldots\gg \rho_d$.
If this is the case, we can substitute the hyperbolic cotangents by
$\pm1$ and get the following equations
\begin{eqnarray}&&
\partial_t\rho_i=\lambda_i+\xi_i+\frac{\kappa}{2}
\exp(-2\rho_i)\,,\quad \lambda_i=\frac{Dd}{2}(d-2i+1)\,.
\nonumber\end{eqnarray}
This simplified dynamics can be easily turned into the probability
density functions of $\rho$ \cite{Kra74,Son98}
\end{multicols}
\begin{eqnarray}&&
{\cal P}(t,\rho_1,\rho_2)=\frac{1}{\sqrt{2\pi Dt}}\exp\left[
-\frac{(\rho_1-Dt)^2}{2D t}
\right]\theta(\rho_1)\delta(\rho_1+\rho_2)\,,
\label{pdfr2}\\&&
{\cal P}(t,\rho_1,\rho_2,\rho_3)
\label{pdfr3}
=\frac{1}{2\sqrt{3}\pi Dt}\exp\left[
-\frac{\left[(\rho_1-3Dt)^2+(\rho_1-3Dt)\rho_2+\rho_2^2\right]}{3Dt}
\right]\theta(\rho_1-\rho_2)\theta(\rho_2-\rho_3)
\delta(\rho_1+\rho_2+\rho_3)\,.
\end{eqnarray}
\begin{multicols}{2}
As explained in the main text these formulae are valid at times
$1/D\ll t\ll 1/D\ln(DL^2/\kappa)$ and the form of the PDF near the boundary
can be modeled by the step function. At times $t\gg 1/D\ln(DL^2/\kappa)$ one
should also calculate the stationary PDF of $\rho_d$. They are readily 
found from the one-dimensional Fokker-Planck equation
\begin{eqnarray}&&
P_{\rm st}(\rho_2)=\frac{\kappa}{4D}
\exp\left(-2\rho_2-\frac{\kappa}{2D}e^{-2\rho_2}
\right)\\&&
P_{\rm st}(\rho_3)=\frac{1}{8\sqrt{\pi}}\left(\frac{\kappa}{D}\right)^{3/2}
\exp\left(-3\rho_3-\frac{\kappa}{4D}e^{-2\rho_3}\right)
\end{eqnarray}

\section{Small $\tau$-expansion}\label{texp}

In this section we assume that the correlation time $\tau$ of $\sigma$ is
small, namely $D\tau\ll 1$, where $D=\langle tr\int dt \sigma^T(0)\sigma(t)
\rangle$ characterizes the amplitude of the fluctuations of $\sigma$.
We investigate the effect of finite correlation
time on the Lyapunov spectrum. In $d=2$ there are no essential changes
with respect to the $\delta$-correlated case since both $\lambda_1$ and
$\lambda_2=-\lambda_1$  get small corrections in $\tau$ leaving all
qualitative features unchanged. However in $d=3$, one can ask whether
$\lambda_2$ will shift from its zero value at $\tau=0$, and whether the
correction is positive or negative at finite $\tau$.
We demonstrate that already the first order correction in $\tau$ leads to
a generally non-zero value of $\lambda_2$ in $d=3$, which can be both
positive and negative. 

Here it is more convenient to parameterize the angular degrees of
freedom of $WW^T$ by the eigenvectors ${\bbox e}_i$
instead of matrix $R$ (see Eqs. (\ref{prt}, \ref{efd})),  
given by $e_i^\alpha=R_{i\alpha}$. If the
eigenvalues are separated, that is $\rho_1\gg\rho_2\gg\rho_3$, the
equations for $\rho_1$ and corresponding to it eigenvector
$\bbox{e_1}$ decouple
\begin{eqnarray}&&
\partial_t{\rho_1}=({\bbox e}_1, \sigma{\bbox e}_1)\,,\quad
\partial_t{\bbox e}_1=\sigma{\bbox e}_1-{\bbox e}_1({\bbox e}_1, 
\sigma{\bbox e}_1)\,.
\end{eqnarray}
The same is true for $\rho_3$ and ${\bbox e}_3$
\begin{eqnarray}&&
\partial_t{\rho_3}=({\bbox e}_3, \sigma{\bbox e}_3)\,,\quad
\partial_t{\bbox e}_3=-\sigma^T{\bbox e}_3+{\bbox e}_3({\bbox e}_3, 
\sigma{\bbox e}_3)\,.
\end{eqnarray}
This system implies that under the transformation
$\sigma\to-\sigma^T$  the eigenvalues are transformed as
$\lambda_{1,3}\to-\lambda_{3,1}$.
In the calculation it is convenient to deal with symmetric matrices, which is 
achieved by decomposing $\sigma$ into symmetric and antisymmetric parts
$\sigma=s+\omega$ and introducing ${\bbox e}_1=M {\bbox n}$ with
\begin{eqnarray}&&
\partial_t M=\omega M\,,\quad M(0)=1
\nonumber\end{eqnarray}
so that
\begin{eqnarray}&&
\lambda_1=\left\langle({\bbox n}, \tilde{s}{\bbox n})\right\rangle\,,\quad
\partial_t{\bbox n}=\tilde{s}{\bbox n}-{\bbox n}({\bbox n}, \tilde{s}{\bbox n})
\nonumber\end{eqnarray}
with $\tilde{s}=M^TsM$. To find the first order correction to $\lambda_1$
we integrate the above differential equation from $0$ to $t$,
and then iterate the obtained expression once. After averaging over the
directions of ${\bbox n}(0)$ we find
\begin{eqnarray}&&
\lambda_1\!=\!\frac{2}{5}\int_0^t \!dt_1\!\left[
tr\left\langle\tilde s(t)\tilde s(t_1)\right\rangle\!+\!
\frac{3}{7}\!\int_0^t \!dt_2\,tr\left\langle {s(t)s(t_1)s(t_2)}\right\rangle
\right],
\nonumber\end{eqnarray}
where $t\gg\tau$. The expression for $-\lambda_3$ is obtained by
changing $s\to-s$ in this formula.
Then, using $\lambda_2=-\lambda_1-\lambda_3$, we find
\begin{eqnarray}&&
\lambda_2=-\frac{12}{35}\int_{0}^t\int_{0}^t\!dt_1dt_2\, 
tr\left\langle {s(t)s(t_1)s(t_2)}
\right\rangle
\nonumber
\end{eqnarray}
which is generally non-zero and has no definite sign.

\section{Pair-correlation function}\label{Pair}
In this appendix we calculate 
the time decay of the pair correlation function
$f(t,r)=\left\langle\vartheta(t,r)\vartheta(t,0)\right\rangle$ within the
Kraichnan model. It satisfies the equation \cite{Kra68}
\begin{eqnarray}&&
\partial_tf={\cal K}_{\alpha\beta}(r)\nabla_\alpha\nabla_\beta f+
2\kappa\nabla^2f\,,\left.\quad f\right|_{t=0}=\chi(r)\,,\nonumber
\\&&
{\cal K}_{\alpha\beta}=D\left(\frac{d+1}{2}\delta_{\alpha\beta}r^2-
r_\alpha r_\beta\right)\,.
\nonumber\end{eqnarray}
Making the Fourier transform over $r$ and passing to spherical coordinates,
we get
\begin{eqnarray}&&
\partial_\tau f=k^2\partial_k^2f+(d+1)k\partial_k f-\epsilon k^2 f
\,,\quad
\label{eq0}
\\&&
\tau=\frac{D(d-1)}{2}t\,,\quad \epsilon=\frac{4\kappa}{D(d-1)}\,.
\nonumber\end{eqnarray}
Next, making the Laplace transform over $\tau$ we obtain
\begin{eqnarray}&&
k^2f''+(d+1)kf'-(E+\epsilon k^2)f=-\chi(k)\,.
\label{eq1}\end{eqnarray}
Two linearly independent solutions of the homogeneous equation are expressed
via the modified Bessel functions  of the order $\nu=\sqrt{E+d^2/4}$. The
branch of the square root should be picked up so that it has a cut along the
semiaxis $E<-d^2/4$ and takes positive values at $E>-d^2/4$. Using these
functions one can find the Green function  $g(E,k,k')$ of Eq. (\ref{eq1})
satisfying the correct boundary conditions
\begin{eqnarray}&&
g(E,k,k')=k^{-d/2}k'^{d/2-1}
\Bigl[
I_\nu(k\epsilon^{1/2}) K_\nu(k'\epsilon^{1/2})\theta(k'-k)\nonumber\\&&+
K_\nu(k\epsilon^{1/2}) I_\nu(k'\epsilon^{1/2})\theta(k-k')
\Bigr],
\label{green}
\end{eqnarray}
with the solution of Eq. (\ref{eq0}) given by
$f(t,k)=\int\! dk'\, g(t,k,k')\chi(k')$.
Next we should make the inverse Laplace transform of $g(E,k,k')$
\begin{eqnarray}&&
g(\tau,k,k')=\frac{1}{2\pi i}\int_{b-i\infty}^{b+i\infty}\! dE 
\, e^{E\tau} g(E,k,k')\,.
\label{inverse}
\end{eqnarray}
Here $b>0$ is arbitrary. One may deform the integration contour in
Eq. (\ref{inverse}) until a singularity of the integrand is encountered.
The first singularity appears at $E=-d^2/4$,
which is the branch point of $\nu$. Therefore, the integration should
be performed along the real axis at $-\infty <E< -d^2/4$ on
both sides of the cut.
Making the change of variable $E=-x^2-d^2/4$ we obtain
\begin{eqnarray}&&
g(\tau,k,k')=\frac{2}{\pi^2} k^{-d/2}k'^{d/2-1}e^{-d^2\tau/4}
\nonumber\\&&
\times\int_0^\infty\!dx\,e^{-x^2\tau} x\sinh(\pi x) K_{ix}(k\epsilon^{1/2})
K_{ix}(k'\epsilon^{1/2})
\nonumber\end{eqnarray}
If one is interested in the single-point statistics,
one should integrate this expression over $k$
\begin{eqnarray}&&
\int \!dk\, k^{d-1}g(\tau,k,k')=\frac{2^{d/2-1}\epsilon^{-d/4}}{\pi^2}
k'^{d/2-1}e^{-d^2\tau/4}\nonumber\\&&
\times\int_0^\infty\!dx\,e^{-x^2\tau} x\sinh(\pi x) 
K_{ix}\left(k'\epsilon^{1/2}\right)
\left|\Gamma\left(\frac{d}{4}+\frac{ix}{2}\right)\right|^2
\nonumber
\end{eqnarray}
Al large $\tau$ the integral is determined by a narrow vicinity of $x=0$.
After a simple calculation we obtain
\begin{eqnarray}&&
\langle\vartheta^2(t)\rangle=\frac{C{\rm Pe}^{d/2}}{(Dt)^{3/2}}
\exp\left[-\frac{d^2(d-1)Dt}{8}\right]
\end{eqnarray}
where $C$ is a $\chi$-dependent constant.

\end{multicols}

\end{document}